# Toward scalable and bias-stable optical phased arrays on lithium tantalate

**Authors:** Gongcheng Yue[1,2][†], Xuqiang Wang[3,4][†], Yihan Miao[2], Bowen Chen[3,4], Yangming Zhan[1], Weiran Zhou[3,5], Phatham loahavilai[2], Jiachen Cai[3,4], Siyuan Yu[1], Chengli Wang[3,4]*, Xin Ou[3,4]*, Yang Li[1]*

**Affiliations:**

[1] State Key Laboratory of Optoelectronic Materials and Technologies, School of Electronics and Information Technology, Sun Yat-sen University, Guangzhou, 510275, China.

[2] State Key Laboratory of Precision Measurement Technology and Instruments, Department of Precision Instrument, Tsinghua University, Beijing, 100084, China.

[3] State Key Laboratory of Materials for Integrated Circuits, Shanghai Institute of Microsystem and Information Technology, Chinese Academy of Sciences, 865 Changning Road, Shanghai 200050, China

[4] Center of Materials Science and Optoelectronics Engineering, University of Chinese Academy of Sciences, Beijing 100049, China

[5] University of Science and Technology of China, 230026 Hefei, China

† These authors contributed equally to this work.

* To whom correspondence should be addressed. Email: wangcl@mail.sim.ac.cn; ouxin@mail.sim.ac.cn; liyang328@mail.sysu.edu.cn;

**Abstract**

Ferroelectric materials are an ideal platform for high-speed reconfigurable photonic integrated circuits (PICs) for classical and quantum photonic computations, communications, and sensing. Most reconfigurable PIC devices achieve their functionalities via interference and are therefore highly sensitive to phase errors. Under static bias, carrier drift in ferroelectric waveguides induces continuous phase drift, creating a severe bottleneck for both PIC functionality and scalability. Here we propose achieving bias-stable and scalable ferroelectric PICs by exploiting the intrinsically low carrier drift of lithium tantalate (LT). Taking one of the PIC devices that is most sensitive to phase drift — the optical phased array (OPA) — as an example, we designed and fabricated an integrated LT OPA that can keep the far-field mainlobe 8 dB higher than sidelobes for over 4 hours, representing at least a two-order-of-magnitude improvement over the state of the art. We demonstrated our device's capability in generating arbitrary spatial-temporal waveforms with a modulation frequency as low as 0.1 Hz, leading to practical applications in optical tweezers, trapped-ion quantum computers, adaptive optics for astronomy, AR, 3D printers, LiDAR, and free-space optical communications. Beyond OPA, our work establishes LT as a bias-stable, scalable, and high-speed PIC platform for large-scale classical and quantum photonic systems.

**Introduction**

Dynamically reconfigurable devices are the essential component of modern photonic integrated circuits (PICs) for classical and quantum communications[1,2], computations[3-5], and sensing[6-8]. Owing to ferroelectric materials' high reconfigurable speed[9-12], they stood out as the workhorse for reconfigurable devices, leading to high-speed micro-ring modulators[13], Mach-Zehnder interferometer-based modulators[14], optical phased arrays[15] (OPAs), and optical neural networks[16] (Fig. 1a). These devices' basic building block is an electro-optic phase modulator (Fig. 1b, left), which changes the phase of the light by altering the ferroelectric waveguide's mode index via the linear electro-optic effect. Ideally, under a static applied voltage, this phase change should be stable over time (Fig. 1c).

However, a ubiquitous physical challenge in ferroelectric waveguides — carrier drift and redistribution — severely disrupts this ideal response[17-19]. Defects in these ferroelectric materials — including oxygen vacancies and impurities — can still create free carriers[20-22]. Under a static applied electric field, these carriers drift along the direction of the electric field (Fig. 1b, right) and are eventually trapped, generating an internal electric field whose direction is opposite to that of the applied electric field[23]. Such an internal electric field increases over time and partially cancels out the applied electric field. Hence, the total electric field over the waveguide gradually decreases, increasing the waveguide's refractive index via the electro-optic effect. Because the carrier drift process is much slower than the electro-optic response, it manifests as a continuous, time-dependent phase drift at the output (Fig. 1c, right).

Crucially, this carrier-induced phase drift is not merely a device-level flaw, it is a fundamental bottleneck that limits the scalability of interference-based PICs. As shown in Fig. 1a, most reconfigurable devices achieve their functionalities via interference, which is sensitive to the phase drift of each phase modulator, especially at larger scales. In systems like OPAs, focusing the far-field beam requires the precise optimization of voltages across a massive phase-modulator array. For large-scale OPAs, the time required to complete this optimization inevitably exceeds the relaxation time of the phase drift, preventing a successful optimization (Fig. 1d, bottom). As the OPA scales down, the total optimization time significantly decreases and can be shorter than the phase-drifting relaxation time, enabling the successful optimization (Fig. 1d, top). Even after this successful optimization, the phase continues to drift over time, resulting in the far-field beam diffusion (Fig. 1e, bottom). Hence, it is imperative to suppress the phase drift to achieve and stabilize the beam focusing, which is the cornerstone of OPA's broad applications (Fig. 1e, top).

Recently, lithium tantalate (LT) has been emerging as a promising material platform for integrated photonics. As a ferroelectric crystal, lithium tantalate features a wide transparent window (0.28−5.5 μm), a moderate refractive index (~2.1@1550nm), a moderate electro-optic coefficient (~30.5 pm/V), and a relatively high Curie temperature ~610 °C. These features, combined with the recently developed LT on insulator (LTOI) wafer fabrication technique and the corresponding nanofabrication technique, open the door to various integrated LT photonic devices and systems[12,24-26]. Compared with lithium niobate (LN) — the ferroelectric material most widely used in integrated electro-optic modulators, lithium tantalate (LT) offers lower birefringence and DC drift, weaker photorefractive effects, and a higher optical-damage threshold[27,28]. Although LN's DC drift can be significantly suppressed by exciting skyrmions in LN, the skymions excitation requires a nontrivial magnetic annealing process[29].

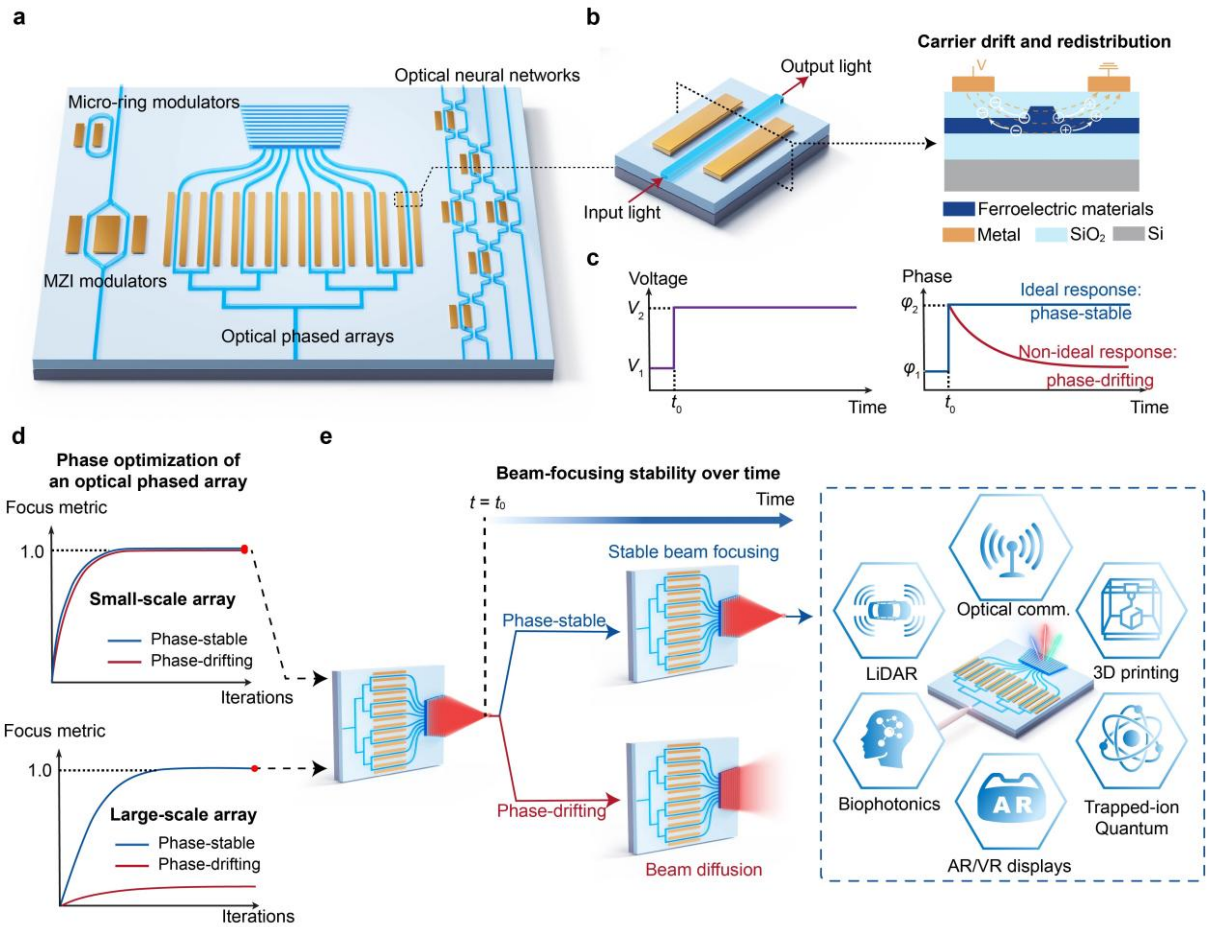

**Figure 1 | Principle of ferroelectric materials-based PIC devices and systems with and without phase drifting. (a)** Schematic of various ferroelectric materials-based PIC devices and systems. **(b)** Phase modulator and its cross-section view illustrating the carrier drift and redistribution. **(c)** Driving voltage (left) and the corresponding ideal and non-ideal phase responses (right) of the phase modulator. **(d)** Phase optimizations for focusing the output beam of small-scale (top) and large-scale (bottom) OPAs. **(e)** Beam focusing performance of OPAs without (top) and with (bottom) phase drifting. Right: OPA's applications in optical tweezers, trapped-ion quantum computers, adaptive optics for astronomy, AR, 3D printers, LiDAR, and free-space optical communications.

Here, we experimentally demonstrate the significant suppression of phase drift in ferroelectric material-based PICs via LT's low DC drift. As a case study, we fabricated and characterized an integrated silicon dioxide-cladded LT OPA that can keep the far-field mainlobe higher than sidelobes up to 16 minutes. To further mitigate the DC drift introduced by silicon dioxide cladding, we realized a cladding-free LT OPA that can keep the far-field mainlobe 8 dB higher than sidelobes for over 4 hours, at least two orders of magnitude longer than that of the state-of-the-art integrated LN OPA[15,30-32]. To reveal the underlying physics of the ultrastable LT PICs, we studied the carrier drift and diffusion process in the LT phase modulator by observing the LT OPA's far-field patterns under different biases. Leveraging the ultra-high stability of the integrated LT OPA, we achieved beam steering with a modulation frequency as low as 0.1 Hz. We also demonstrated the arbitrary spatial-temporal waveform generation, enabling OPA's practical applications in optical tweezers[33], trapped-ion quantum computers[34,35], adaptive optics for astronomy[36], AR[37], 3D printers[38], LiDAR[39], and free-space optical communications[15] (Fig. 1e). The proposed LT platform also enables various highly scalable reconfigurable PIC devices and systems, such as large-scale classical and quantum computing systems.

**Scalability of ferroelectric materials-based OPAs**

To quantitatively elucidate the impact of phase drift on device scalability, we performed numerical simulations of the phase optimization process of OPAs with different array sizes using a stochastic parallel gradient descent algorithm. In the optimization of the phases of the OPA's phase modulator array, the optimization target was the Strehl ratio — the ratio of the optimized far-field mainlobe intensity to its ideal theoretical value. As shown in Fig. 2a, the Strehl ratio of the ideal 16-channel OPA converges to 1 after ~150 iterations. After similar iterations, the Strehl ratio of the 16-channel OPA, whose phase drifts 0.1 radian per iteration, can converge to ~0.9. In contrast, for a 128-channel OPA with the same phase drift, the Strehl ratio does not converge even after 4,000 iterations. Without such a phase drift, the Strehl ratio can converge to 1 after 1,300 iterations. These results confirm that phase drift critically influences the scalability of OPAs.

To further quantify the effect of phase drifting on OPA's scalability, we simulated OPAs with and without phase drifting at different scales. As shown in Fig. 2b, as the channel number increases from 15 to 110, the iteration number that the ideal OPA needs to reach 0.5 Strehl ratio linearly increases from 67 to 360. In contrast, the iteration number of the OPA with phase drifting exponentially increases as a function of the channel number, reaching 4,500 at the 110 channel number. When the channel number is larger than 120, the Strehl ratio can not converge to 0.5 anymore. Because practical applications such as LiDAR and free-space optical communications strictly require large-scale arrays to synthesize narrow beamwidths, these results confirm that mitigating phase drift is an absolute prerequisite for scaling ferroelectric PICs.

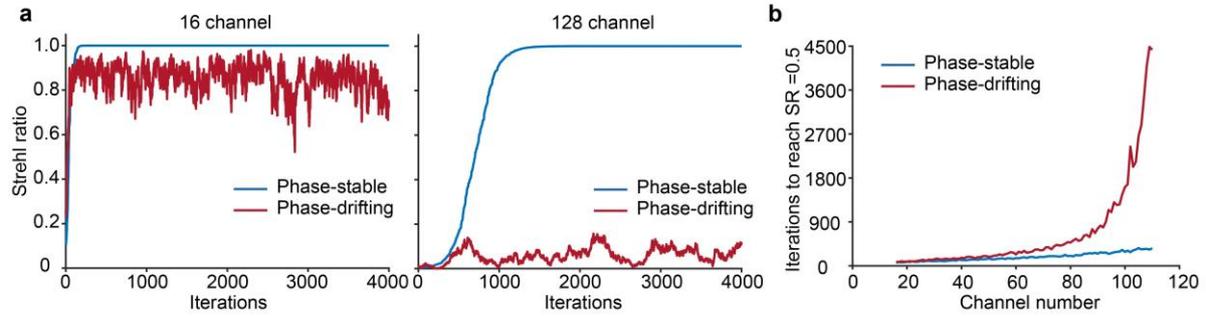

**Figure 2 | Scalability analysis of ferroelectric materials-based OPAs. (a)** The ratio of optimized mainlobe intensity to its ideal value (Strehl ratio) as a function of optimization iteration of 16-channel OPAs with and without phase drifting. **(b)** Same as (a) for 128-channel OPAs. **(c)** The iteration number that is required to reach the Strehl ratio of 0.5 as a function of the channel number of OPAs with and without phase drifting.

**Device fabrication and characterization**

DC drifting is the phenomenon that a ferroelectric electro-optic phase modulator's output light's phase gradually drifts over time under a static driving voltage (Fig. 3a). DC drifting originates from the carriers' migration and redistribution over the ferroelectric waveguide under the applied electric field. These carriers induce an internal electric field that is opposite to the applied electric field and, in turn, at least partially cancels the applied electric field, leading to a smaller total electric field over the waveguide. This smaller total electric field increases the waveguide's mode index via the electro-optic effect, resulting in a phase drift of the output light of the phase modulator.

We propose to reduce ferroelectric material-based OPA's phase drift by leveraging LT's weaker DC-drifting effect compared with LN (Fig. 3a). We attribute LT's weaker DC-drifting effect to the following two reasons. Compared with LN, LT has a wider bandgap[40], reducing the probability of generating photon-induced carriers for building up the internal electric field. On the other hand, LN's defects, such as anti-site niobium ions, can provide the trapping centers for carriers, facilitating the redistribution of carriers and the build-up of the internal electric field[41]. And, the stronger internal electric field better cancels the applied field, enhancing the DC-drifting effect. Anti-site niobate ions' counterpart in LT — anti-site tantalate ions have different energy levels and trapping/releasing dynamics[42], reducing or slowing down the accumulation of carriers and the build-up of the internal field, and eventually weakening the DC-drifting effect.

We propose further reducing the integrated LT OPA's phase drift by removing the silicon dioxide cladding (Fig. 3a). The interface between the ferroelectric waveguide and silicon dioxide cladding provides additional donor and trapping energy levels, facilitating carriers' migration and trapping near the interface[43-45]. Hence, the transport path of carriers can be reduced by removing the silicon dioxide cladding, weakening the build-up of the internal electric field. And, a weaker internal electric field eventually reduces the DC-drifting and the corresponding phase drift.

We fabricated and packaged a 16-channel integrated LT OPA using the standard planar process (Figs. 3b, c). We coupled a continuous-wave (CW) laser into the device via lensed fiber. Then, we equally split the CW laser into 16 channels via cascaded multimode interferometers. We then input each channel's CW laser into a phase modulator with a 1.5-cm modulation length. Finally, to decrease the channel spacing and, in turn, increase the OPA's field of view, we combined the 16 channels into a narrow area to input into a slab grating.

We measured the far-field pattern and half-wave voltage of the LT OPA using the experimental setup shown in Fig. 3e. We used an objective lens to Fourier-transform the OPA's near field to the far field, which is imaged by the camera via two lenses. The imaged far-field pattern shows diffused spots along the phase-modulation direction due to the phase errors originating from the differences between channels' lengths (Fig. 3g top, 3h left). We calibrated such phase errors using the rotating element vector algorithm (See Supplementary information S3), resulting in a well-converged far-field beam spot (Fig. 3g bottom, 3h right). This beam spot exhibits full-width-of-half-maximum (FWHM) beam widths of 2.6° along the phase-modulation direction and 1.6° along the wavelength-modulation direction, respectively. After calibration, we measured the half-wave voltage of each channel by varying the phase of a specific channel while keeping the phases of the other channels constant, depicting a half-wave voltage of 4.4 V (Fig. 3f).

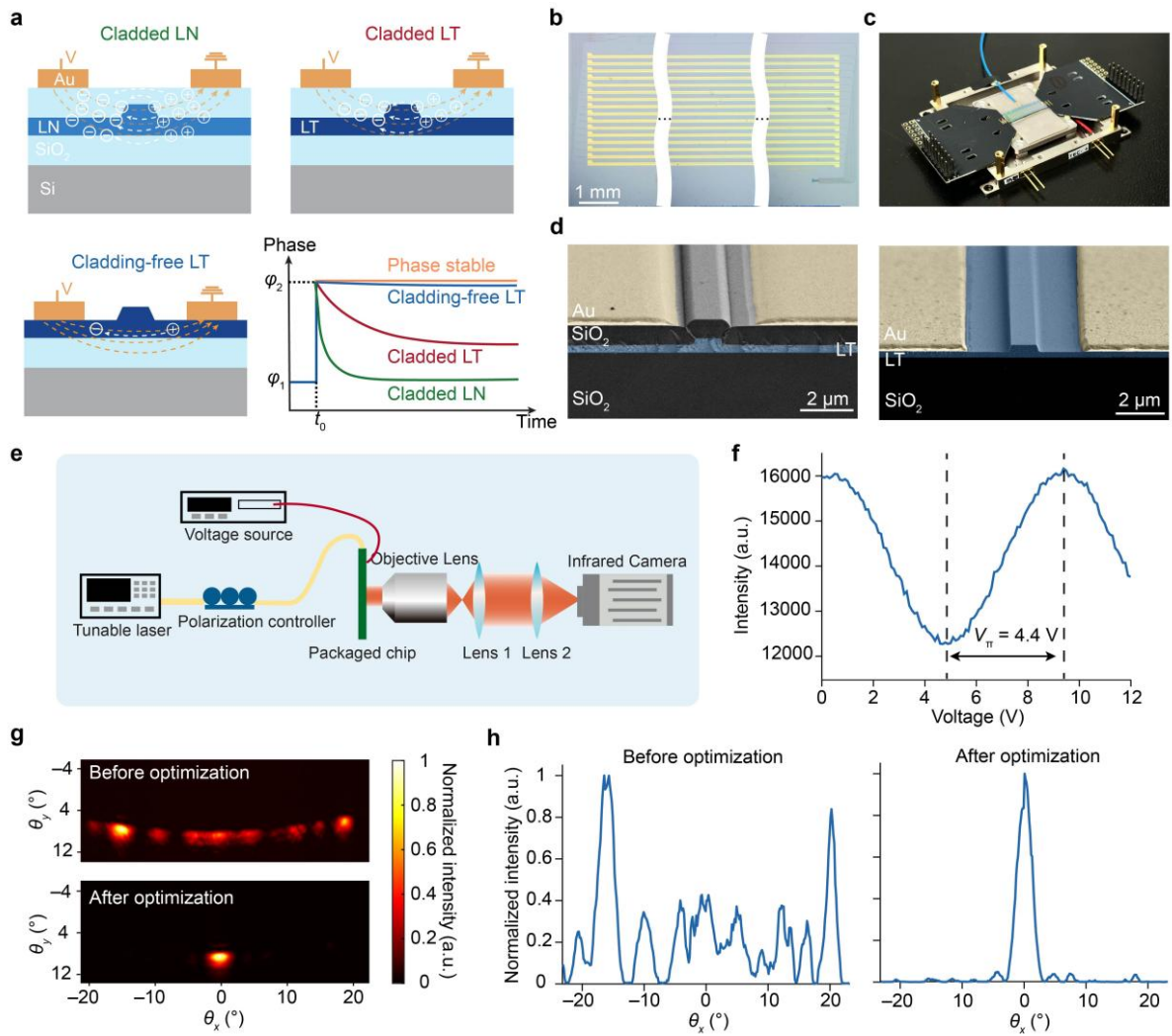

**Figure 3 | Integrated LT electro-optic OPA's fabrication and measurement results**. **(a)** Schematics of the cross-sections of LN cladded, LT cladded, and LT cladding-free phase modulators, leading to distinct phase responses over time (bottom right). **(b)** Optical microscope image of the fabricated device. **(c)** Photo of the device after packaging. **(d)** Scanning electron microscope (SEM) image of LT cladded (left) and LT cladding-free (right) devices. **(e)** Experimental setup for far-field measurement. **(f)** Normalized optical transmission as a function of the applied voltage. **(g)** Measured far-field patterns of the OPA before (top) and after (bottom) optimization. **(h)** Cross sections of the measured far-field patterns before (left)

and after (right) optimization.

**Silicon dioxide-cladded LT OPA**

We fabricated and measured the far-field patterns of a silicon dioxide-cladded LT OPA (Fig. 3d, left). As shown in Figs. 4a, c, d, e, the sidelobe levels increase over time under a constant bias, resulting in a lower SLSR. We define the focusing-stability time as the time at which the SLSR degrades to 0 dB, which is ~16 min for this device (Fig. 4a). Owing to the power redistribution from the main lobe to sidelobes, as the SLSR degrades, the mainlobe intensity decreases, reaching half (−3 dB) of its maximum after ~7 minutes (Fig. 4a).

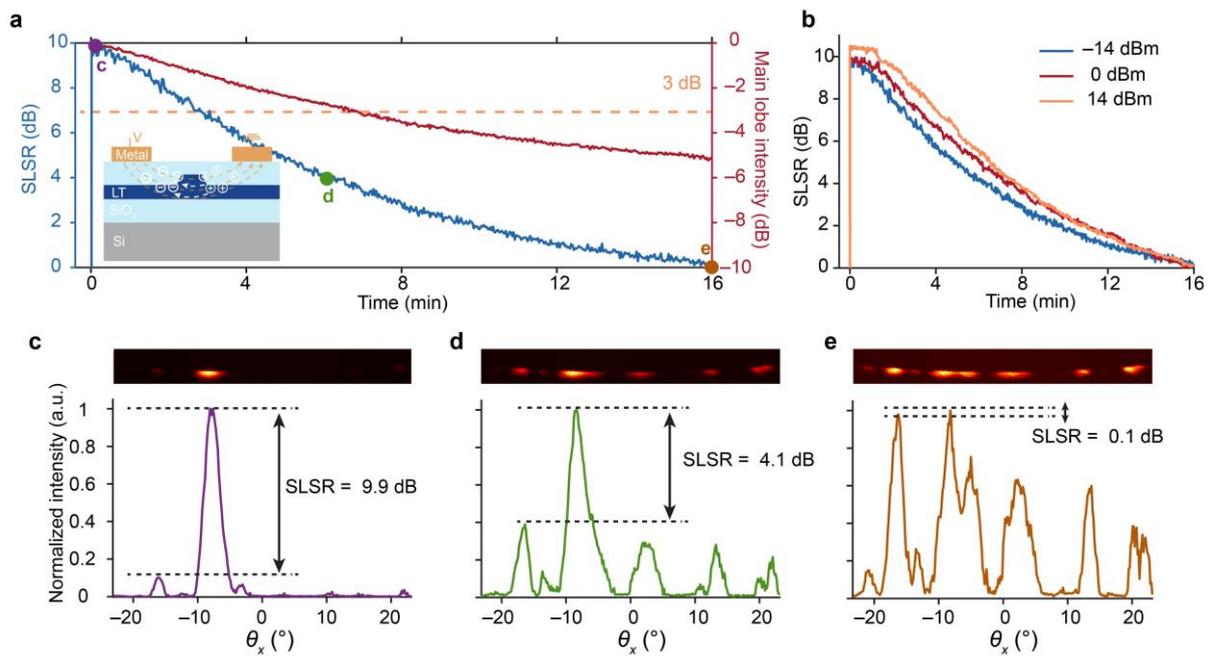

**Figure 4 | Silicon dioxide-cladded integrated LT electro-optic OPA's schematic and measurement results. (a)** Measured SLSR (left) and main lobe intensity (right) over time. Inset: schematic of carrier drift across the LT waveguide under a given voltage. **(b)** Measured SLSRs under optical powers of 14 dBm, 0 dBm, and −14 dBm. Measured far-field patterns (top) and their cross-sections (bottom) at $t = 0$ **(c)**, 6 minutes **(d)**, and 16 minutes **(e)** in **(a)**.

We attribute the focusing-degradation behavior to the different phase-drift rates of all the LT phase shifters under distinct voltages. For each phase shifter, the carrier in the LT and silicon dioxide drift under a fixed voltage, forming an internal electric field. This internal electric field alters the LT waveguide's effective refractive index via the electro-optic effect, leading to the phase drift in each channel. Such a phase drift's speed — the phase drift rate — differs across different LT phase shifters because the optimized voltages are distinct from one phase shifter to another, leading to different phase changes to distinct channels. This non-uniform phase-drifting phenomenon induces the focusing degradation.

To test the cladded LT OPA's focusing stability under different optical powers, we measured SLSRs at three distinct optical powers. As shown in Fig. 4b, SLSRs under optical powers of 14 dBm, 0 dBm, and −14 dBm show very similar behavior — decreasing over time and approaching zero near 16 minutes. This weak power dependence suggests that photorefractive contributions, which are typically proportional to optical intensity, are not the dominant origin of the observed SLSR degradation.

To investigate carrier drift in the cladded LT waveguides, we measured the far-field patterns over time (Fig. 5). As shown in Fig. 5b, before applying the bias to the phase shifters, the far-field pattern shows randomly distributed beams, corresponding to the equilibrium carrier distribution across each waveguide and a random phase distribution over the channels. Once the biases are applied to the phase shifters, each phase shifter immediately reaches its ideal phase via the linear electro-optic effect, leading to the well-defined far-field beam at $\theta_x \approx -8°$ (Fig. 5a (i)). Under the bias-induced constant external electric field, the carriers gradually drift across each waveguide over time, forming an internal electric field and, in turn, altering each waveguide's effective index via the electro-optic effect. Hence, each channel's phase gradually deviates from its ideal value, leading to several sidelobes (Fig. 5b (ii)). After 4 hours, the internal electric field almost cancels out the bias-induced external electric field (Fig. 5c, right), leading to a far-field pattern (Fig. 5c, left) similar to that before applying the bias (Figs. 5b).

We also study the carrier diffusion in the cladded LT waveguides. As shown in Fig. 5d (right), once the bias is removed, the internal electric field with the same amplitude but opposite direction relative to the bias-induced external electric field results in a moderately-defined far-field beam at $\theta_x \approx 10°$ (Fig. 5d, left), which is almost a mirror symmetry of the well-defined far-field beam at the moment that the bias is applied (Fig. 5a (i)). After removing the bias, the carriers gradually diffuse over the waveguide, redistributing the power from the moderately-defined far-field beam to the sidelobes (Fig. 5a (iii)). After removing bias for more than 1 hour, the carrier diffusion process reaches a steady state, leading to the equilibrium carrier distribution across the waveguide and a far-field pattern (Fig. 5a (iv)) very similar to that before applying the bias (Fig. 5b).

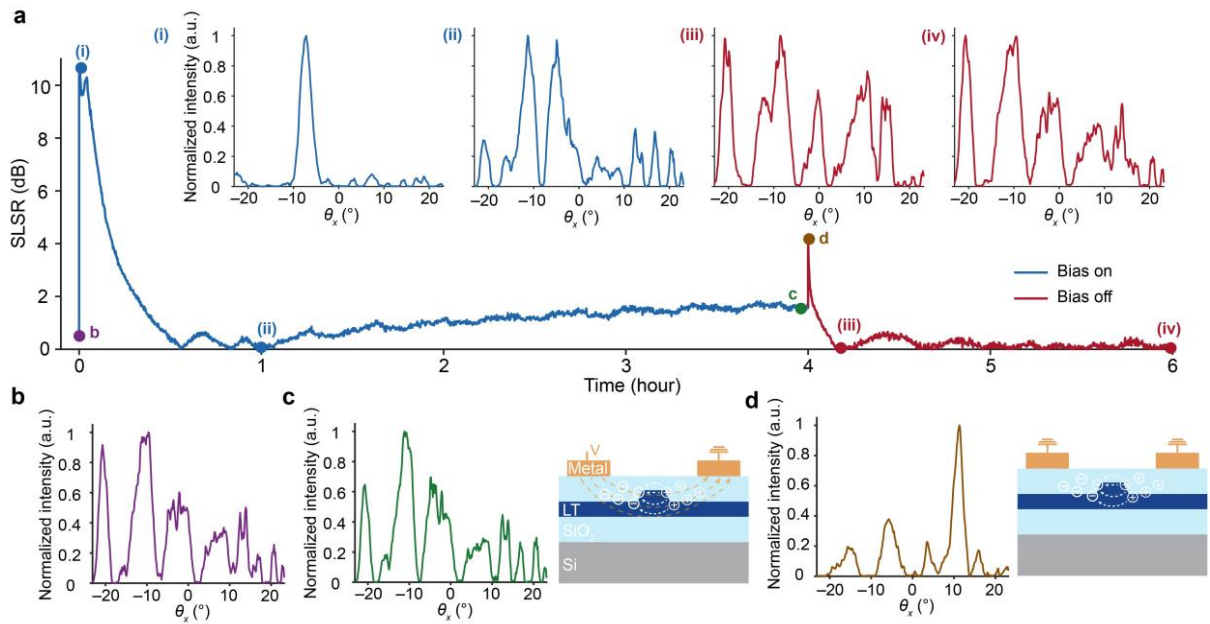

**Figure 5 | Silicon dioxide-cladded integrated LT electro-optic OPA's measurement results.** **(a)** Measured SLSR over time with bias on and off. Insets show measured far-field patterns' cross sections at the time of (i) 0 second, (ii) 1 hour, (iii) 4 hours 15 minutes, and (iv) 6 hours. **(b)** Measured far-field pattern's cross section before applying the bias. **(c)** Measured far-field pattern's cross section before switching off the bias (left). The corresponding carrier distribution over the LT waveguide (right). **(d)** Measured far-field pattern's cross section after switching off the bias (left). The corresponding carrier distribution over the LT waveguide (right).

**Cladding-free LT OPA**

To further improve the focusing stability, we fabricated and measured a cladding-free LT OPA (Fig. 3d, right). As shown in Fig. 6a, this device shows an SLSR over 8 dB for over 4 hours. The corresponding focusing-stability time of over 4 hours is at least two orders of magnitude longer than that of cladded LT OPA (Figs. 6a(i)-(iii)). Such a significant improvement in the focusing-stability time originates from the absence of the silicon dioxide cladding, which enhances the carrier migration and trapping[43,46].

As a control experiment, we fabricated and tested a 16-channel cladded LN OPA with a similar design. Although this LN OPA's small scale requires a short optimization time for focusing the far-field beam in a specific direction, we still can not achieve this focusing because this LN film's phase-drift relaxation time is shorter than the optimization time (Fig. 6a). If the LN OPA is fabricated using different recipes, the far-field beam's focusing could be optimized and maintained for a few minutes[30-32], corresponding to focusing-stability time of a few minutes. Because this focusing-stability time is even shorter than that of our cladded LT OPA, we can conclude that our cladding-free LT OPA's focusing-stability time is at least two orders of magnitude longer than the state-of-the-art integrated LN OPA.

To test the cladding-free LT OPA's focusing stability, we measured the SLSR for three distinct optical powers. As shown in Fig. 6b, the SLSRs corresponding to 14 dBm, 0 dBm, −14 dBm are closely assemble each other, demonstrating that the photorefractive effect has an even weaker influence on cladding-free OPA when compared with that on cladded OPA (Fig. 4b).

We tested the beam steering performance of the cladding-free LT OPA. As shown in Fig. 6c, by adjusting the phases of the 16 phase shifters, we can sweep the optical beam from −20° to 20°, covering an angular range of 40° along the $\theta_x$ direction. As shown in Fig. 6d, we can steer the optical beam from 15° to 6° by sweeping the CW laser wavelength from 1520 nm to 1620 nm, depicting an angular range of 9° along the $\theta_y$ direction. Leveraging the cladding-free LT OPA's phase modulation, wavelength modulation, and high focusing stability, we demonstrated 2D beam steering with a low modulation frequency down to 1 Hz and a high modulation frequency up to 100 Hz (see Movies S1, S2).

To showcase the integrated LT OPA's capability in arbitrary optical beam manipulation, we measured the far-field patterns as a function of time. As shown in Fig. 6e, we manipulated the optical beam by varying the voltages applied to the phase shifters, leading to the trajectory showing the triangular-, square-, and sinusoidal waveforms as a function of $\theta_x$ and time. These waveforms feature a temporal period as long as 10 s, corresponding to a modulation frequency as low as 0.1 Hz. This result demonstrates our device's capability in fixing the optical beam for a long time and manipulating the optical beam at a high speed, enabling broad applications such as optical tweezers, AR, and free-space optical communications.

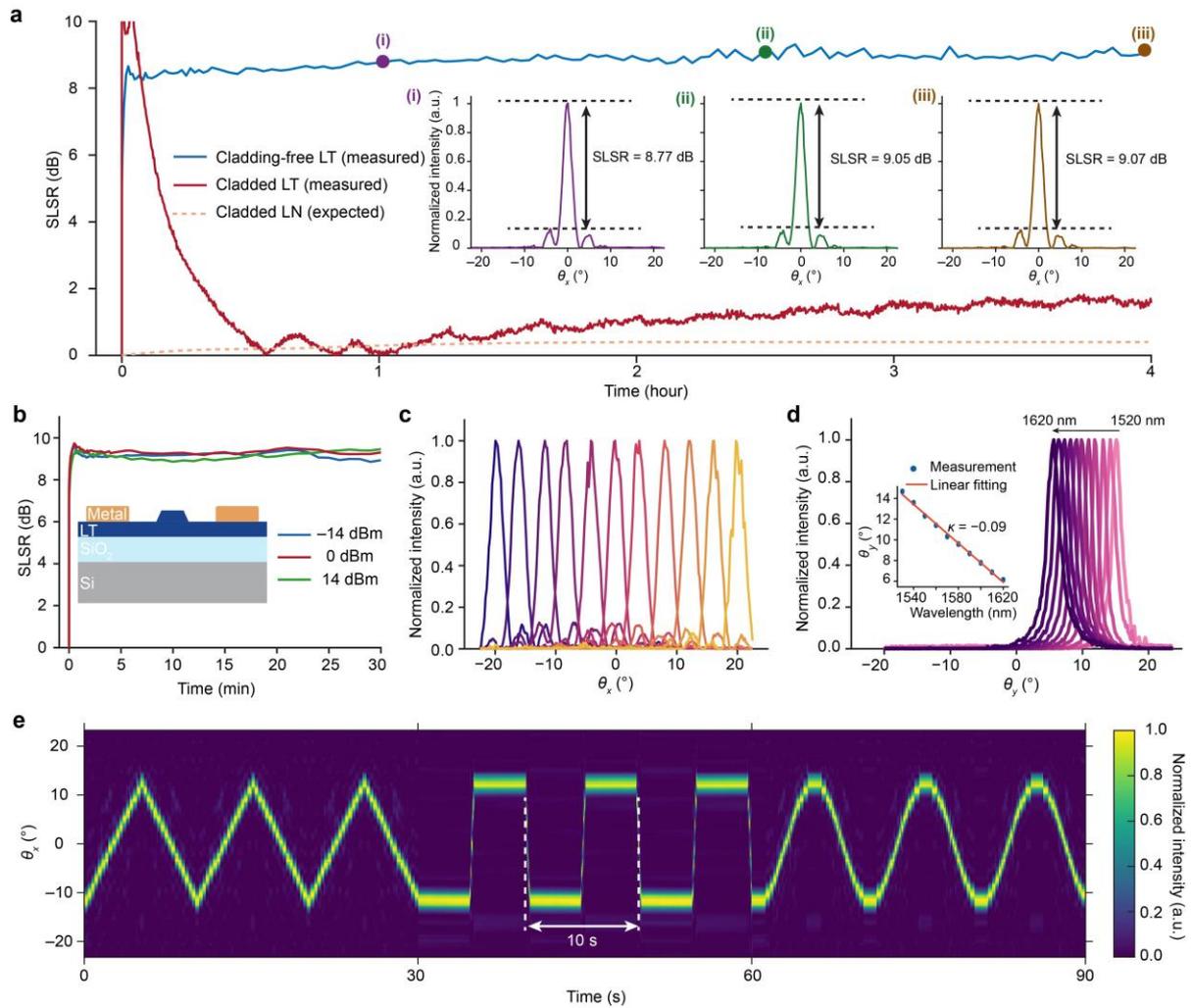

**Figure 6 | Stability and beam manipulation of integrated LT electro-optic OPAs. (a)** Measured SLSRs of integrated cladding-free LT and cladded LT OPAs, as well as expected SLSRs of integrated cladded LN OPA over time. Insets show the cross-sections of the measured far-field patterns at $t$ = 1 hour (i), 2.5 hours (ii), and 4 hours (iii). **(b)-(d)** Measurement results of the integrated cladding-free LT OPA. **(b)** Measured SLSRs under optical powers of 14 dBm, 0 dBm, and −14 dBm. **(c)** Measured far-field patterns over the $\theta_x$ cross-section. **(d)** Measured far-field patterns over the $\theta_y$ cross-section. The inset shows the measured and linearly fitted $\theta_y$ from 1520 nm to 1620 nm. **(e)** Triangular-, square-, and sinusoidal spatial-temporal waveforms generated by the integrated LT OPA.

**Conclusion and discussion**

We significantly suppressed ferroelectric material-based PICs' phase drift via LT's low DC

drift, enabling high-speed bias-stable PIC systems at large scales. As a case study, we demonstrated an integrated silicon dioxide-cladded LT OPA that can keep the far-field mainlobe higher than sidelobes up to 16 minutes. To further improve the focusing-stability time, we demonstrated a cladding-free integrated LT OPA that can keep the far-field mainlobe 8 dB higher than sidelobes for over 4 hours, which is at least two orders of magnitude longer than that of the state-of-the-art integrated LN OPA. Leveraging this high stability, we achieved beam steering with a modulation frequency as low as 0.1 Hz. We also demonstrated our device's capability in arbitrary spatial-temporal waveform generation, opening the door to exploring integrated electro-optic OPA's applications in optical tweezers, trapped-ion quantum computers, adaptive optics for astronomy, AR, 3D printers, LiDAR, and free-space optical communications. Beyond OPA, the bias-stable and scalable LT PICs also unlock large-scale high-performance computing, communicating, and sensing systems.

Looking forward, the evolution of bias-stable LT PICs presents several exciting avenues for future research. Crucially, our cladding-free demonstration successfully isolates the intrinsic phase-stable LT platform from the extrinsic carrier trap dynamics typically introduced by conventional silicon dioxide claddings. Having established this fundamental baseline for ultrastability, future work can focus on co-optimizing the LT platform with advanced, low-drift passivation materials. This will seamlessly combine the demonstrated pristine phase stability with the robust environmental protection required for commercial foundry packaging. Furthermore, meeting the stringent density requirements of next-generation PICs will necessitate scaling down device footprints. This can be addressed by leveraging advanced modulator architectures, such as transparent metal oxide electrodes[47], dual-layer electrodes[48], silica-LT hybrid waveguide[49], and waveguide with isolation trenches[50]. Ultimately, the functional scope of this LT platform can be maximized through heterogeneous integration[26]. By seamlessly interfacing large-scale LT optical neural networks with active components like semiconductor lasers and germanium photodetectors, we envision the realization of fully monolithic, bias-stable photonic computing and sensing systems.

**Methods**

*Fabrication process*

The integrated LT OPA was fabricated on an x-cut TFLT-on-insulator wafer using a standard planar process. We prepared the wafer from an optical-grade LT bulk wafer and converted it into a thin-film-on-insulator stack by ion cutting and wafer bonding process (Smart-cut technique). The stack consisted of a 600-nm-thick LT membrane, a 4.7-μm-thick buried $SiO_2$ layer, and a 500-μm-thick silicon substrate. We diced the wafer into 2 cm×1.5 cm chips to allow for more flexible device fabrication.

We patterned the waveguide using e-beam lithography (Elionix ELS-BODEN 125) with hard mask (Ma-N 2410), followed by argon-ion-based physical dry etching (Leuven, HAASRODE-I200). We then used an alkaline wet etching process to remove the sidewall redeposition and residual resist. After waveguide fabrication, we annealed the chip at 500 °C for one hour in a nitrogen atmosphere to repair etch-induced damage. We then deposited a $SiO_2$ cladding layer using inductively coupled plasma chemical vapour deposition (SENTECHSI 500D) to form high-quality $SiO_2$ upper-cladding. Finally, we fabricated the modulation electrodes by a dual-layer lift-off process based on maskless lithography (Heidelberg DWL 66+) and an electron-beam metal evaporation system (DENTON EXPLORER-14). The electrodes consisted of a 300-nm-thick Au layer on a 10-nm-thick Ti adhesion layer.

After completing chip fabrication, we packaged the integrated LT OPA to provide stable optical and electrical interfaces for subsequent characterization. We aligned and fixed a commercial polarization-maintaining lensed fiber to the edge couplers of the photonic integrated circuit. We then mounted the fiber-pigtailed chip on a custom evaluation printed circuit board (PCB) to enable wire bonding, and used gold wire bonds to connect the on-chip pads to the corresponding metal pads on the PCB.

*Details of the experimental setup*

Schematic diagrams of the experimental setups are shown in Fig. 3e. We used a continuous-wave (CW) laser (Santec TSL-550) as the light source. We set the input polarization to TE using a polarization controller (JCOPTIX FAP-C56) before coupling the light into the chip.

After the light passes through the chip and exits from the slab grating, an objective lens (NA = 0.4), with the slab grating located at its front focal plane, collects the emitted light. For near-field imaging, we removed lens 1 and positioned the infrared camera (Hamamatsu, 14041-10U) at the back focal plane of lens 2. In this configuration, the objective and lens 2 image the near field of the slab grating onto the infrared camera. To prevent overexposure, we placed several neutral density filters (JCOPTIX OFA14) in front of the camera. We used the near-field image to align the optical path and locate the slab grating, ensuring efficient collection of the grating emission at the camera. After alignment, we inserted lens 1 back into the optical path and aligned its front focal plane with the back focal plane of the objective. The objective lens then performed a Fourier transform of the near field into the far field, and lenses 1 and 2 imaged this far-field pattern onto the infrared camera.

*Data and code availability*

The data and code used in this study are available from the corresponding authors upon request.

**Acknowledgements**

This work received support from the National Key Research and Development Program of China (2023YFB3211200 and 2025YFA1411704), the National Natural Science Foundation of China (62435009 and 62293521), the Strategic Priority Research Program of Chinese Academy of Sciences (XDB1440200), and the Shanghai Science and Technology Innovation Action Plan Program (24CL2901000 and 24CL2901002). The sample fabrication is supported by SIMIT material-device and characterization platform and ShanghaiTech Material and Device Lab (SMDL). The authors thank Lu Ye and Xiaofeng Wang from SJTU-Pinghu Institute of Intelligent Optoelectronics for packaging devices. The authors thank Rongjin Zhuang and Jinze He for the helpful discussion on the methodology.

## Author contributions

Y.L., X.O., G.C.Y., and X.Q.W. conceived the basic idea for the work. G.C.Y and X.Q.W designed the device. X.Q.W., B.W.C., W.R.Z., and J.C.C. fabricated and characterized the chip. G.C.Y., Y.H.M., Y.M.Z. and P. L. carried out the device measurement. G.C.Y. and X.Q.W. analyzed the measurement results. G.C.Y. prepared the figures. C.L.W., X.O., and Y.L. supervised the research. G.C.Y. and Y.L. wrote the original draft of the manuscript and all the authors reviewed and contributed to the manuscript.

## Competing interest declaration

The authors declare no competing financial interests.

## Additional information

*Supplementary Information* is available for this paper.

*Corresponding author*

Correspondence and requests for materials should be addressed to C.L.W. (wangcl@mail.sim.ac.cn), X.O. (ouxin@mail.sim.ac.cn) and Y.L. ([liyang328@mail.sysu.edu.cn](liyang328@mail.sysu.edu.cn)).